# Measurement of quantum states of neutrons in the Earth's gravitational field

V. V. Nesvizhevsky,[1,*] H. G. Börner,[1] A. M. Gagarski,[2] A. K. Petoukhov,[1,2] G. A. Petrov,[2] H. Abele,[3] S. Baeßler,[3,7] G. Divkovic,[3] F. J. Rueß,[3] Th. Stöferle,[3] A. Westphal,[3] A. V. Strelkov,[4] K. V. Protasov,[5] and A. Yu. Voronin[6,5]

[1]*ILL, Grenoble, France*
[2]*PNPI, Gatchina, Russia*
[3]*University of Heidelberg, Germany*
[4]*JINR, Dubna, Russia*
[5]*ISN, Grenoble, France*
[6]*FIAN, Moscow, Russia*
[7]*University of Mainz, Germany*

The lowest stationary quantum state of neutrons in the Earth's gravitational field is identified in the measurement of neutron transmission between a horizontal mirror on the bottom and an absorber/scatterer on top. Such an assembly is not transparent for neutrons if the absorber height is smaller than the ''height'' of the lowest quantum state.

## I. INTRODUCTION

Quantum stationary states of matter in electromagnetic and strong fields are well known. Examples are eigenstates of electrons in atoms, and the ones of protons and neutrons in nuclei. It is straightforward to expect the existence of quantum states of matter in a gravitational field as well. In the laboratory the gravitational field alone does not create a potential well, as it can only confine particles by forcing them to fall along gravity field lines. We need a second ''wall'' to create the well. The analytical solution of the Schrödinger equation, which describes the quantum states of a particle with mass $m$ above a horizontal reflecting mirror, is known and can be found in textbooks [1–3]. However, up to the present experiment no such quantum effect had been observed. It is very difficult to measure because gravitational quantum effects are negligible for macroscopic objects and the electromagnetic interaction dominates for charged particles. Therefore one should use neutral elementary particles with a long lifetime. Such a scenario with slow neutrons was proposed in Ref. [4] and later discussed in Ref. [5]. Recently a high precision neutron gravitational spectrometer has been built and installed at the Institute Laue-Langevin. It is optimized for this scenario and related studies [6]. We will discuss here an experiment, which allowed to observe for the first time the lowest quantum state of neutrons in the Earth's gravitational field. The results of this experiment were summarized in Ref. [7]. Here we will give a detailed account of all experimental aspects concerning this work. Additional discussions, analytical solutions, and related publications can be found elsewhere, for instance, in Refs. [8–22].

## II. BASIC PRINCIPLES

In the quasiclassical approximation the value of the $n$th energy level $E_n$ $(n=1,2,3,...)$ is calculated using the Bohr-Sommerfeld formula:

$$E_n=\sqrt[3]{\left(\frac{9\cdot m}{8}\right)\cdot\left[\pi\cdot\hbar\cdot g\cdot\left(n-\frac{1}{4}\right)\right]^2}. \qquad (1)$$

This approximation works extremely well for linear potentials and provides almost an exact solution even for the ground state $(n=1)$. Only acceleration in the Earth's gravitational field $g$, the neutron mass $m=940\ \mathrm{MeV}/c^2$, and the Planck constant $\hbar=6.61\times10^{-16}$ eV/s define the energy values $E_n$. The energies of the four lowest levels are 1.4 peV, 2.5 peV, 3.3 peV, and 4.1 peV, respectively. The spatial neutron density distribution, which corresponds to the lowest levels, is shown in Fig. 1. The probability $\psi_n^2(z)$ of finding neutrons at the height $z$ has $n$ maxima and $(n-1)$ minima with the probability zero in each minimum as for any ordinary standing wave.

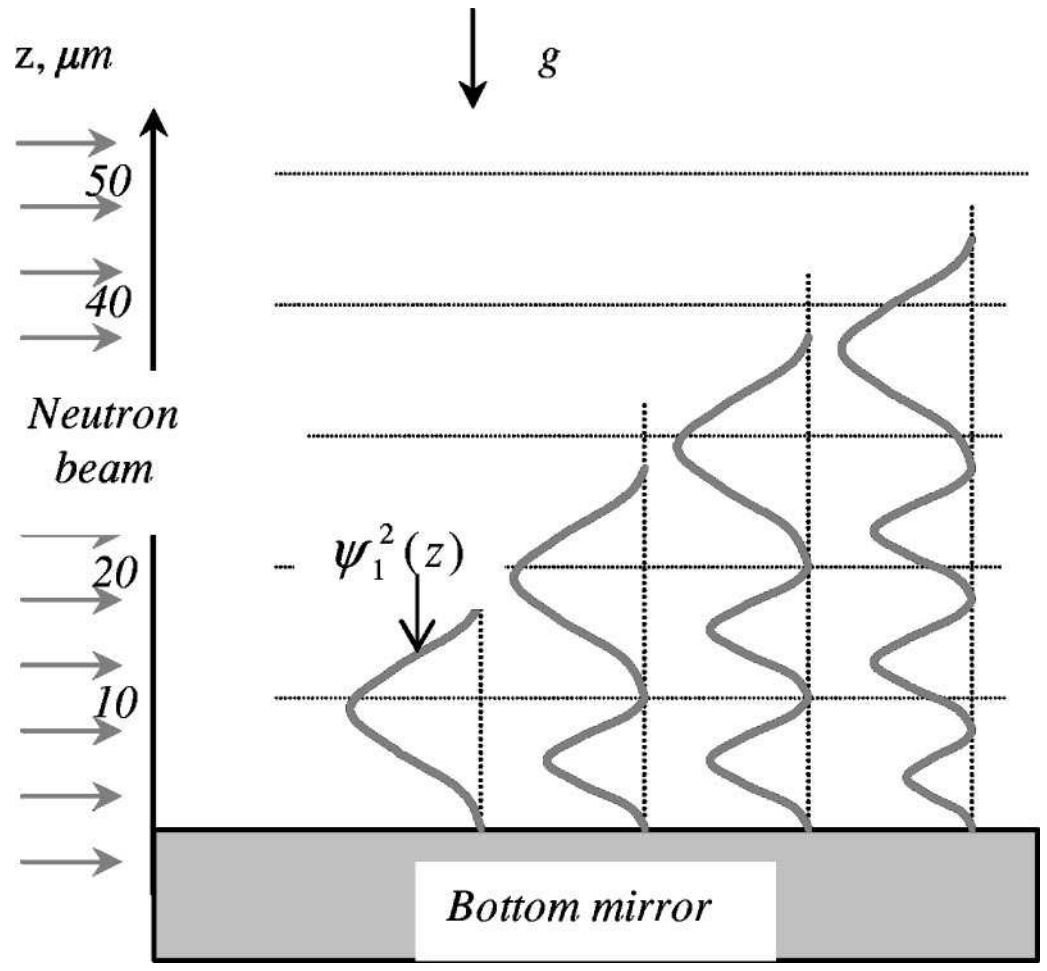

FIG. 1. Quantum states of neutrons are formed in the potential well between the Earth's gravitational field on top and a horizontal mirror on bottom. The probability of finding neutrons at a height $z$, corresponding to the $n$th quantum state, is proportional to the square of the neutron wave function $\psi_n^2(z)$. The vertical axis $z$ gives an idea about the spatial scale for this phenomenon.

*Corresponding author. Email address: nesvizhevsky@ill.fr

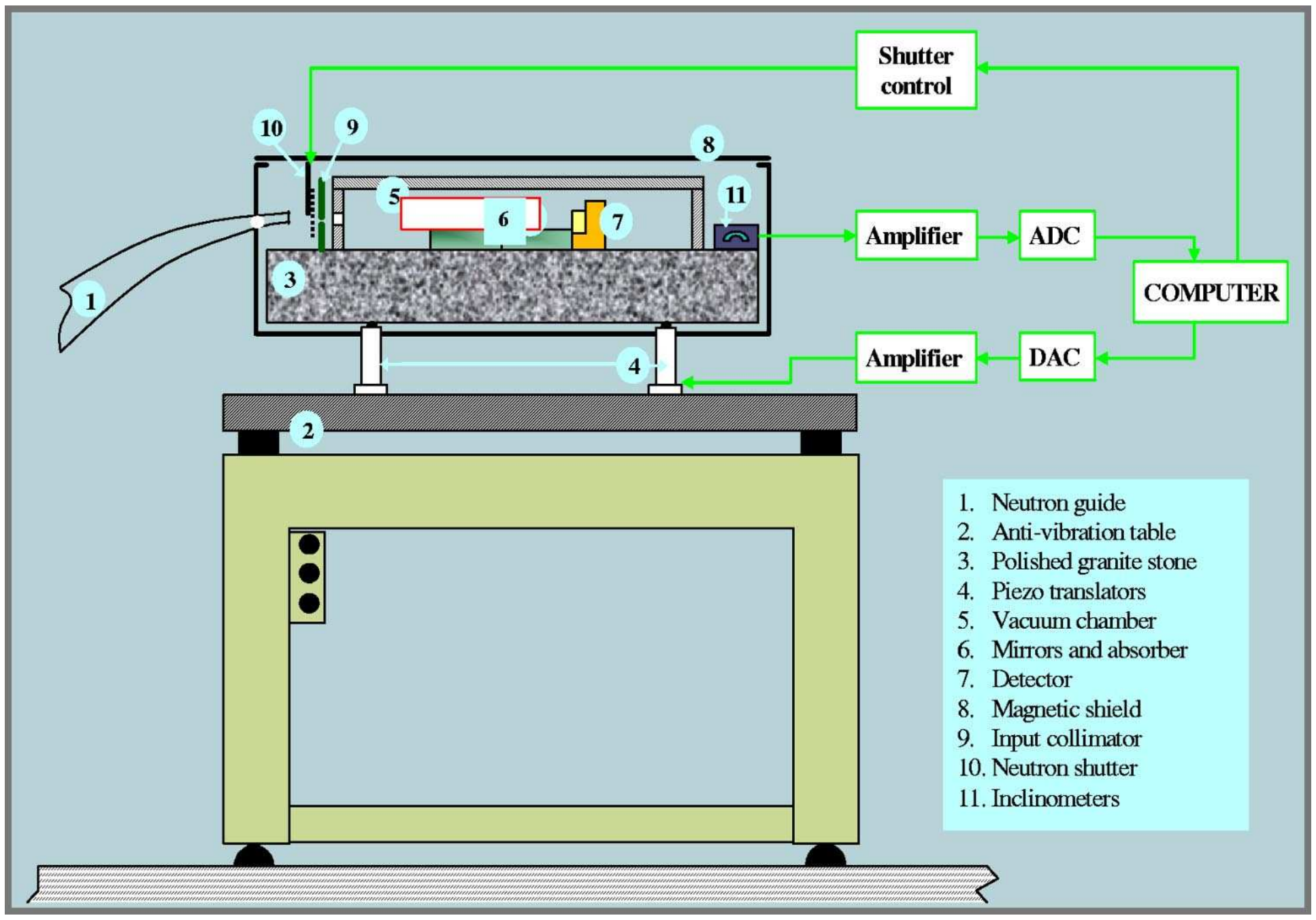

FIG. 2. The experimental setup. The gravitational levels are expected to be formed above the bottom mirrors (6). The mirrors and the detector (7) are placed inside a vacuum chamber (5), which has a 30-$\mu$m-thick Al entrance window. Between the exit of the neutron guide (1) and the vacuum chamber entrance (separated by a distance of $\sim$1.5 cm) an adjustable input collimator (9) is installed. The separation between the neutron guide and the experimental setup allows to mechanically isolate the setup against parasitic vibrations, originating from the neutron guide. The complete setup is mounted on a standard pressurized antivibration table (2). A system of three interconnected pneumatic valves provides leveling of the ‘‘floating’’ tabletop. A polished flat granite slab (3) is mounted to the ‘‘floating’’ table top on three active legs which consist of piezotranslators (4). They are connected in a closed loop with precise inclinometers (11), installed on the granite surface. Length variation of the piezotranslators enables automatic leveling of the granite with an absolute precision better than 10 $\mu$rad. A magnetic shield (8) insures screening of external magnetic field gradients. The entrance neutron guide (1) ends in a piece which has at its exit a horizontal opening of 13 cm length and of 1 cm height. A thin Al foil of thickness 30 $\mu$m is glued to its exit. The entrance collimator (9) has two thick horizontal titanium plates with length and height larger than the size of the exit window of the neutron guide. The height position of every plate can be adjusted independently with the accuracy of $\sim$10 $\mu$m. This titanium collimator, in combination with the absorber and the bottom mirror (6), allows for selection of neutron velocity components (see Fig. 4 below).

The basic idea of this experiment is to let neutrons ‘‘flow’’—with a certain horizontal velocity—between a mirror below and an absorber/scatterer above. The absorber acts as a selector for the vertical velocity component. Then one measures the number $N$ of transmitted neutrons as a function of the absorber height $\Delta h$. This measurement should allow the identification of the neutron quantum states because a clear ‘‘signature’’ should appear: the classical dependence $N_{\mathrm{CL}}(\Delta h)$ is modified into a stepwise quantum-mechanical dependence $N_{\mathrm{QM}}(\Delta h)$ at small absorber heights $\Delta h$. This is described below.

## III. EXPERIMENTAL SETUP

The experimental setup is shown in Fig. 2. Details are described in the text and in Ref. [6].

Neutrons of broad velocity and angular distribution arrive at the entrance of the setup. The spectrum of the horizontal velocity components of neutrons was measured. The results are described below and shown in Fig. 3. The angular distribution of neutrons at the entrance to the mirrors/absorber is illustrated in Fig. 4. The vertical angular divergence ($\sim 4 \times 10^{-2}$) is defined by relative positions of titanium collimating plates and the front edge of the bottom mirror. The arrangement of the absorber and the mirror reduces the range of incoming vertical velocities by an order of magnitude. The neutron transmission between the bottom mirror and the absorber was measured as a function of the inclination angle of the complete setup. This showed that there is no correlation and demonstrated high uniformity of the phase space density of incoming neutrons. The upper cutoff in the vertical velocity component is defined by the absorber height. Only neutrons that can penetrate (between the bottom mirrors and the absorber) are those which have a very small vertical velocity component at the front edge of the bottom mirror (see Fig. 4). This allows a simple determination of the horizontal velocity component. The lower cutoff in the horizontal velocity component defines a parabolic trajectory to start at the edge of the lower titanium collimator plate. The higher cutoff is done by the edge of the higher collimator plate. In actual measurements the heights of these two titanium plates were adjusted in such a way that there was no other cutoff in the spectrum of horizontal velocity components. This spectrum is shown in Fig. 3. The horizontal velocity component of 5–15 m/s is much bigger than the vertical velocity of $< 5$ cm/s. Therefore, in a vertical slice, one can find several neu-tron "planar jets" with a density dependency on the distance from the bottom mirror(s).

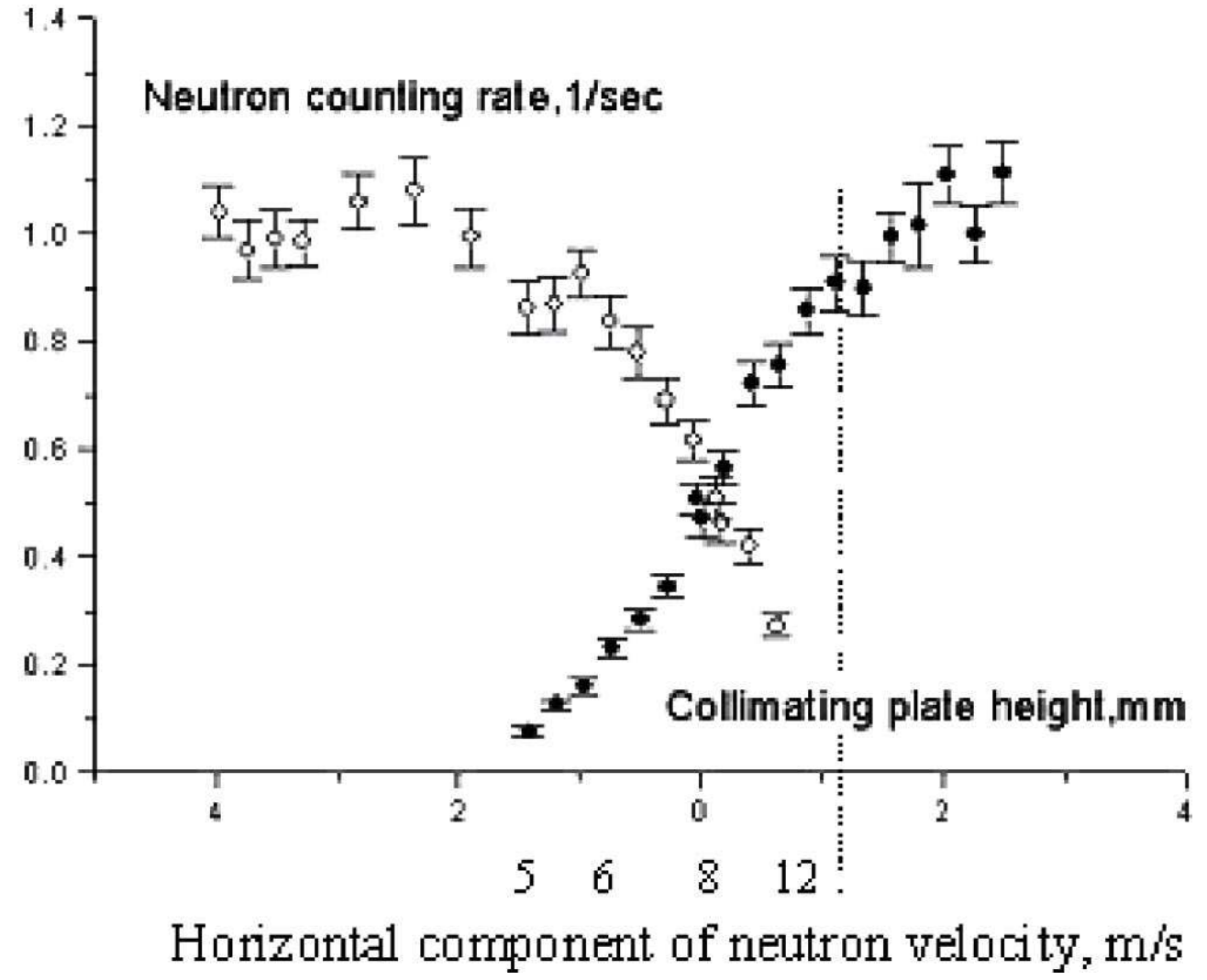

FIG. 3. The spectrum is shown of the horizontal velocity components perpendicular to the detector surface. The solid points correspond to the height variation of the upper collimator plate (the lower plate has the minimal height). The open circles correspond to the height variation of the lower collimator plate (the upper plate has the maximal height). The zero height corresponds to the half intensity in the neutron flux measured with the absorber at the height of 150 $\mu$m. The neutron horizontal velocity component is shown. It is calculated using the length of $\sim$12 cm of parabolic trajectories.

The phase space neutron density, which corresponds to the lowest quantum states, is low. The count rate was $10^{-3}$–$10^{-1}$ $s^{-1}$ in the "quantum" measuring regime (low absorber height, see Fig. 5). On the other hand, the total neutron flux through the entrance neutron guide is $\sim 10^6$ $s^{-1}$, which is $10^7$–$10^9$ times higher! Therefore efficient background suppression is important in such a measurement. The background caused by external thermal neutrons was suppressed by "4-$\pi$ shielding" of the detector with $B_4C$ rubber. A long narrow opening in the shielding of only a few millimeters in height allows to count those neutrons which penetrate between the bottom mirror(s) and the absorber. Particular attention was paid to the part of the neutron background which results from multiple elastic reflection of ultracold neutrons (UCN) inside the experimental setup. A multislit shielding [Fig. 4(e)] between the entrance Al window and the mirrors protects the detector also against those UCN which have been up-scattered (in energy) in the neutron guide or in the entrance collimator. However, this multislit shielding does not influence the spectrum of vertical and horizontal components of velocity of neutrons counted in the detector. There is no direct view from the neutron guide to the detector while the neutrons, which are then trapped in the quantum levels states, move into the setup via curved parabolic trajectories. The upper titanium plate inhibits the direct view to the detector.

The bottom mirrors are standard optical polished glass plates. Their surfaces were studied with small angle scattering of x rays [24]. The root mean square height roughness is $\sim$10–20 Å. The lateral correlation length of the surface roughness is $\sim$10 $\mu$m. Such a surface is sufficiently flat in order to provide specular reflections of neutrons from the mirror such that parallel and normal to surface motions of neutrons are rather independent. This provides a continuous neutron spectrum in the horizontal plane, but quantum states for the vertical motion.

The absorber/scatterer is a glass plate with high macroscopic flatness and with high microroughness. The roughness is a few micrometers. The correlation length of roughness is a few times larger. This glass plate is then coated on one side (using the magnetron evaporation method) with a Ti-Zr-Gd alloy (in the proportion 54%-11%-35%, 0.2 $\mu$m thickness). This is usually used as an antireflecting sublayer in polarizers for cold neutrons [23]. Neutrons can be absorbed in it or scattered from the absorber surface, mainly nonspecularly.

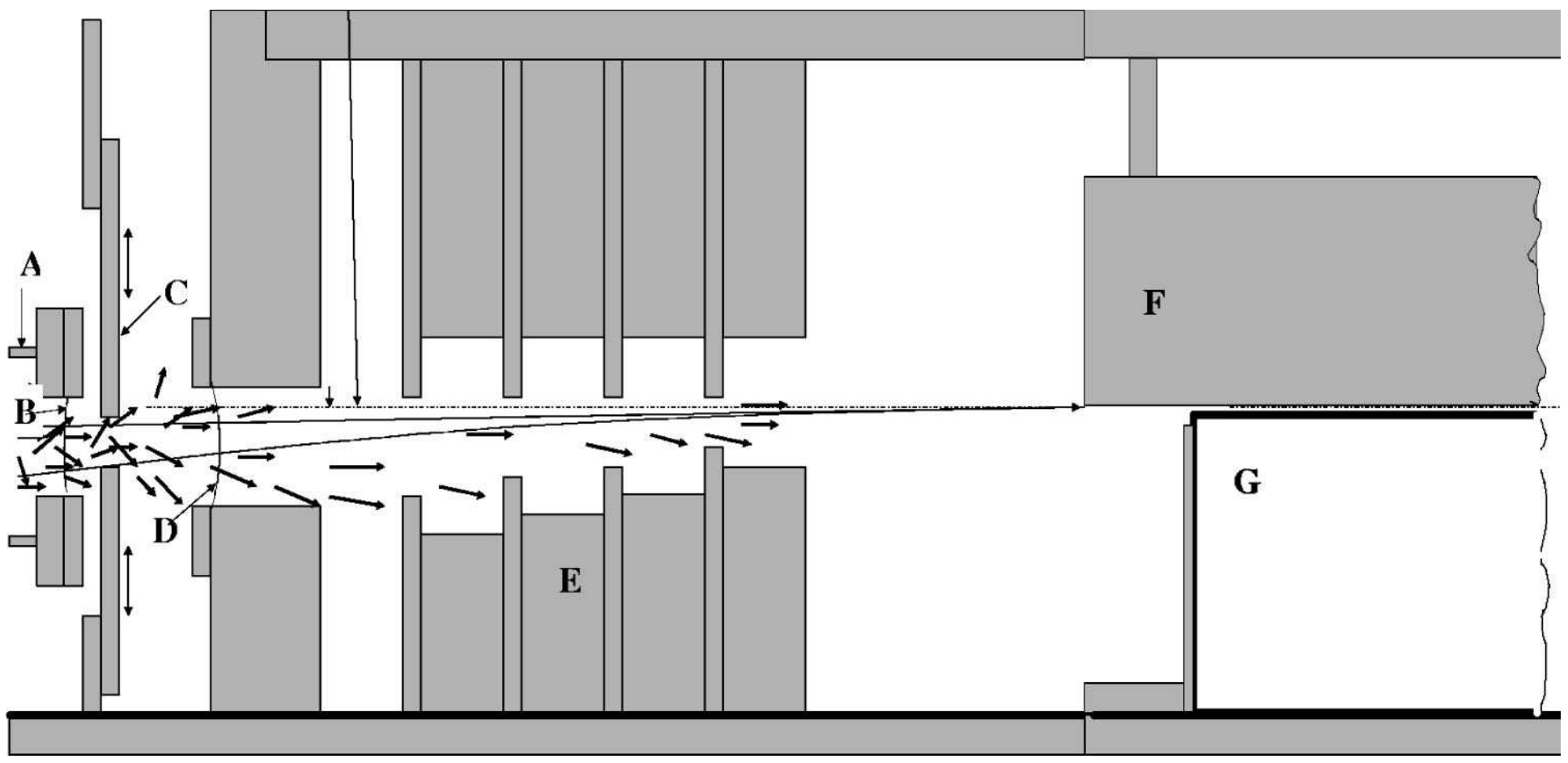

FIG. 4. A vertical cut through the collimators and a scheme of selection of neutron velocities and trajectories are shown. Absorbing materials are marked in gray. Long arrows illustrate those trajectories of neutrons, which can then penetrate into the detector. Short arrows illustrate "lost" trajectories. At the entrance side the absorber exceeds the bottom mirror by 3–5 cm. (A) Neutron guide, (B) its exit window, (C) titanium collimators, (D) entrance window, (E) multislit shielding, (F) absorber, and (G) bottom mirror.

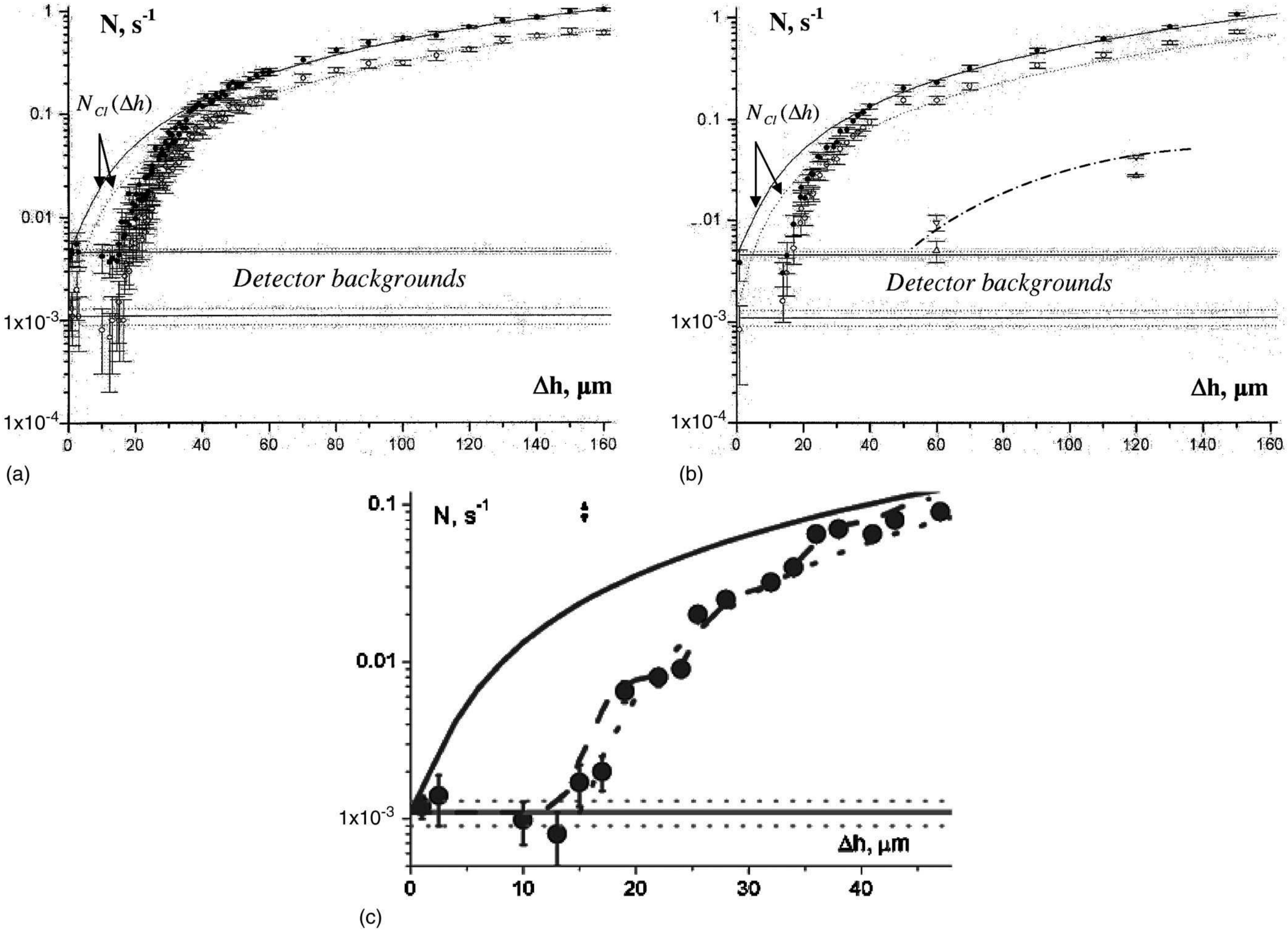

FIG. 5. (a) The dependence $N(\Delta h)$ for two bottom mirrors with the length of 6 cm. The solid circles show the data measured in the detector discrimination window "counting of all events." The solid curve fits these data with the classical dependence. The open circles correspond to the data measured in the "peak" discrimination window. The dotted curve fits these data with the classical dependence. The horizontal straight lines show the detector background values and their uncertainties measured with the reactor "off." The present data were measured in four sets. In every set the height was varied from zero to the maximum value. (b) The dependence $N(\Delta h)$ for the bottom mirror with the length of 10 cm. The data and the fits are analogous to those shown in (a). The much lower (recall the logarithmic scale) triangles show the data measured with inverse geometry: the absorber on the bottom and a 10 cm mirror on top. The dash-dotted line corresponds to a dependence $N(\Delta h)$ calculated for inverse geometry. (c) The summed dependence $N(\Delta h)$. Four measurements with two 6 cm mirrors, two other measurements after complete readjustment of the setup, and three measurements with a 10 cm bottom mirror are included. The data are summed up in intervals of 2 $\mu$m. The solid circles correspond to the data measured in the discrimination window "counting of all events." The open circles correspond to the data measured in the "peak" discrimination window. The stepwise broad curves correspond to the fits using $N_{\mathrm{QM}}(\Delta h)$ dependence with idealized energy resolution. The dotted curve corresponds to $N_{\mathrm{QM},0}(\Delta h)$ dependence for the discrimination window "counting of all events." The solid smooth curve corresponds to the classical dependence $N_{\mathrm{CL}}(\Delta h)$ for the same data.

Such a nonspecular reflection mixes horizontal and vertical velocity components and increases strongly the frequency of their collisions with the absorber. This mechanism allows to reject neutrons with a vertical velocity component, which is high enough to reach the absorber.

A classical estimation for the distance $\Delta l$ between two subsequent collisions of a neutron with the bottom mirror (if it does not touch the absorber) is

$$\Delta l_{\mathrm{CL}} = 2\cdot V_{\mathrm{hor}}\cdot\sqrt{\frac{2\cdot h_0}{g}}, \tag{2}$$

where $h_0$ is the maximal elevation of a neutron in the gravitational field and $V_{\mathrm{hor}}$ is the horizontal velocity component.

For our neutron spectrum, if the mirror and the absorber are longer than ~10 cm then neutrons cannot pass through (in the classical approximation), without touching the bottom mirror and/or the absorber at any actually used absorber height. This condition allows to estimate the minimal mirror length which is sufficient for shaping the spectrum of the vertical velocity components. The uncertainty principle provides an even shorter value for the minimal mirror length needed to resolve different quantum states (Refs. [5–7]):

$$\Delta l_{QM} = V_{hor} \cdot \frac{\hbar}{(E_2 - E_1)} \tag{3}$$

(for the two lowest quantum states). The mirror should be significantly longer than these estimations [Eqs. (2) and (3)].

We measured the neutron transmission between the bottom mirror and the absorber as a function of the absorber height. The absorber is mounted on three active legs. The length of each leg can be varied to within 250 $\mu$m by means of a piezoelement. A precise two-axis inclinometer measures the relative shift of the absorber position when the legs are lifted subsequently one after another. The neutron count rate $N(\Delta h)$ is measured in a cylindrical $^3$He gaseous detector (2.3 bars Ar, 30 torr $^3$He, 10 torr $CO_2$, $\sim$20 cm long, $\sim$1.7 cm in diameter). The entrance window for neutrons (12 cm long and 1.5 mm high) is made of 100 $\mu$m Al foil. The detector is installed inside the vacuum chamber. It operates without electric discharges if the residual gas pressure is lower than $\sim 2\times10^{-2}$ torr. The Al entrance window of the detector would reflect neutrons with a normal velocity component lower than 3.2 m/s, but the arrangement of the separating Al windows (the exit window of the neutron guide [(1), Fig. 2] and the entrance window of the vacuum chamber [(5), Fig. 2] had already removed such neutrons from the spectrum.

## IV. THE MEASUREMENT OF THE NEUTRON TRANSMISSION

A measurement of the count rate of neutrons penetrating between the bottom mirror and the absorber versus the absorber height allowed identification of the quantum neutron states.

Two discrimination windows in the pulse height spectrum of the $^4$He detector were set as follows: (1) A "peak" discrimination window corresponded to the narrow peak of the reaction

$$n + {}^3\text{He} \rightarrow t + p, \quad Q = 0.764 \text{ MeV} \tag{4}$$

and provided low background; (2) a much broader range of amplitudes ($E > 0.15$ MeV) allowed "counting of all events." Usually the detector discrimination levels were adjusted such that $\sim$50% of the neutron events were counted in the first discrimination mode and $\sim$90% were counted in the second discrimination mode. The background was $(1.3 \pm 0.2)\times10^{-3}$ s$^{-1}$ and $(4.6\pm0.3)\times10^{-3}$ s$^{-1}$, respectively, when the neutron reactor was "off." Figure 5 demonstrates that the background was very efficiently suppressed: when the absorber height was zero and the neutron reactor was "on" then the count rate corresponded to within statistical accuracy the detector background measured with the neutron reactor "off." Other background sources, such as thermal neutrons from the reactor hall and UCN elastically or inelastically scattered inside the setup, were negligible, as can be seen in Fig. 5.

Two configurations of the bottom mirror(s) and the absorber were used. The width of two identical bottom mirrors in the first measurement was 10 cm and the length was 6 cm. These two mirrors were installed behind each other, with a slit between them smaller than about 5 $\mu$m. The height of these two mirrors was equal to within a few micrometers. The absorber width was 10 cm; its length was 13 cm. The absorber was shifted towards the entrance so that it did not cover the last 2 cm of the bottom mirror on the detector side. In the second measurement one bottom mirror of identical thickness and width and of 10 cm length was installed instead of the two shorter bottom mirrors.

In an "ideal" experiment, described bellow, if $\Delta h$ is smaller than, roughly speaking, the "height" of the lowest quantum state $h_1$ then $N_{QM}$ should be zero. When $\Delta h \approx h_1$ then $N_{QM}$ should increase sharply. Further increase in $\Delta h$ should not increase $N_{QM}$ as long as $\Delta h$ is smaller than the "height" of the second quantum state $h_2$. Then, at $\Delta h \approx h_2$, again $N_{QM}$ should increase stepwise. At sufficiently high $\Delta h$ one should approach, for uniform distribution of the phase space neutron density, the classical dependence of

$$N_{QM}(\Delta h) \rightarrow N_{CL}(\Delta h) \sim (\Delta h)^{1.5} \tag{5}$$

and the stepwise increase should be washed out. The power factor 1.5 in the classical dependence is due to simultaneous increase in the slit size (power factor 1) and in the range of acceptable vertical velocity components (power factor 0.5). The later one is a consequence of the gravitational field only. This is only precisely true if all neutrons are lost with a vertical velocity component, which is sufficient to raise in the gravitational field up to the absorber height. A comparison of the experimental data to this asymptotic dependence provides a good test for the quality of the absorber. Figures 5(a) and 5(b) show that the classical dependence $N_{CL}(\Delta h)$ fits the data well at large $\Delta h$ values.

On the other hand, when we put a second reflecting mirror on top (instead of the absorber), we expect to get the dependence

$$N_{test}(\Delta h) \sim \Delta h, \tag{6}$$

which was also well confirmed in the corresponding measurement with two mirrors.

In the measurement with the mirror on bottom and the absorber on top, we found that at lower heights the count rate $N(\Delta h)$ is considerably suppressed compared to the classical dependence $N_{CL}(\Delta h)$. This is expected from quantum mechanics. The strongest effect is measured when the absorber height is smaller than the "height" $h_1$ of the lowest quantum state of neutrons in the gravitational field (see Fig. 1). Such an assembly is not transparent at all. Neglecting the "fine" stepwise structure of the dependence $N_{QM}(\Delta h)$, one could approximate it as

$$N_{QM,0} \sim (\Delta h - h_1)^{1.5}. \tag{7}$$

Here $h_1 \sim 15$ $\mu$m corresponds to the energy of the lowest quantum state of 1.4 peV. This shift in height takes into account the effective reduction of available phase space compared to the classical model due to the uncertainty principle.

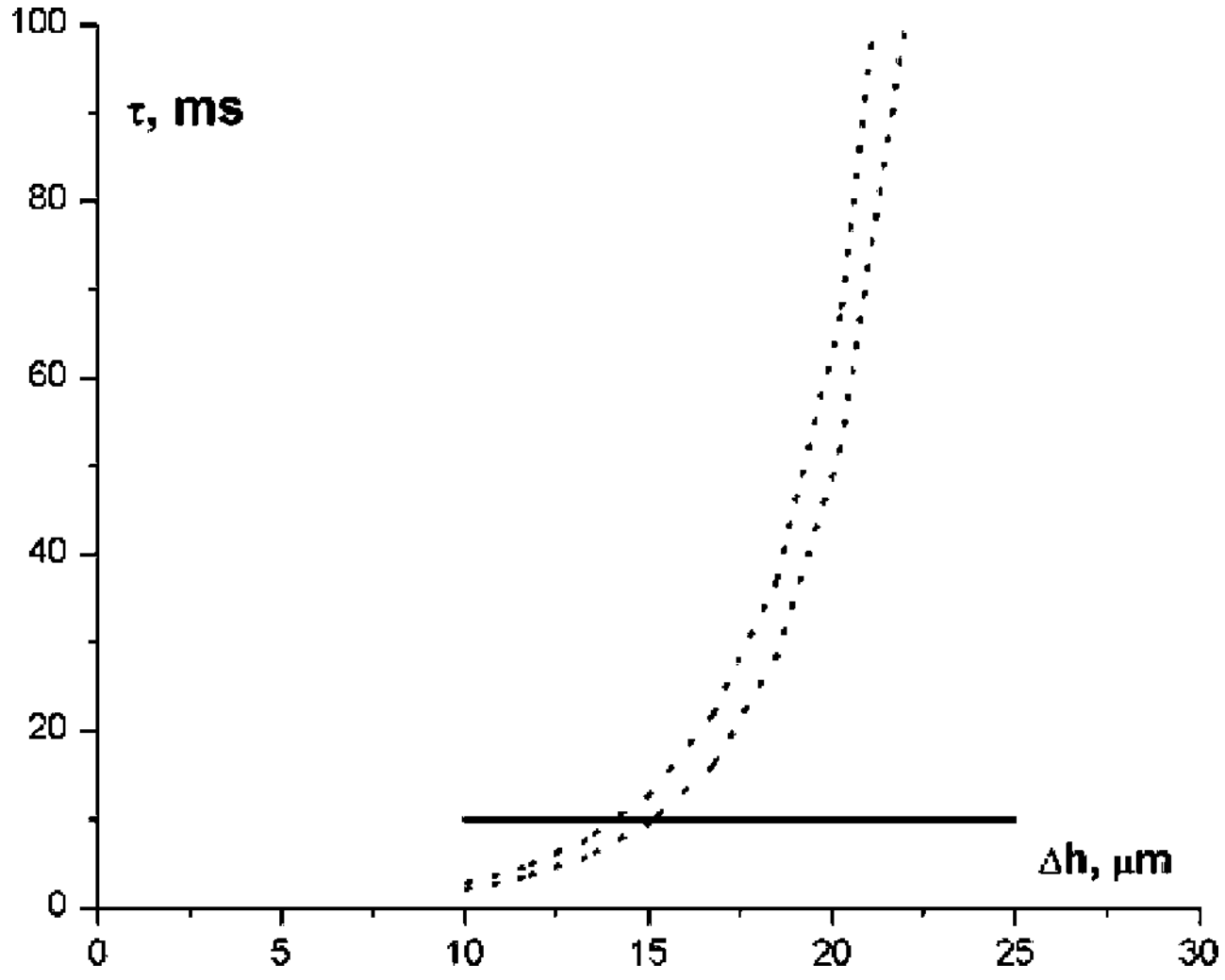

FIG. 6. Characteristic cleaning time for neutrons in the lowest quantum state is shown as a function of the absorber height. The absorber strength for these two curves differs by a factor of 3. Horizontal line corresponds to the average time of flight of neutrons through the mirror/absorber slit.

Figure 5(c) shows that the dependence $N_{\mathrm{QM},0}(\Delta h)$ corresponds approximately to the general behavior of the experimental data points.

A more adequate and simple quantum-mechanical treatment $N_{\mathrm{QM}}(\Delta h)$ is based on quasiclassical Eq. (1). Let us assume that (i) the absorber captures all neutrons at the levels with the energies higher than the absorber's "potential" energy $m\cdot g\cdot\Delta h$. Figure 6 shows that the neighbor quantum levels can be separated. The dependence on the absorber efficiency is weak. (ii) The absorber does not shift significantly the energy of other "opened" levels. Figure 7 shows that the energy of the lowest level is almost constant if the absorber is higher than 15 $\mu$m, and also the dependence on the absorber strength is weak. These two hypotheses were verified by quantum-mechanical calculations with a complex potential reproducing our absorber.

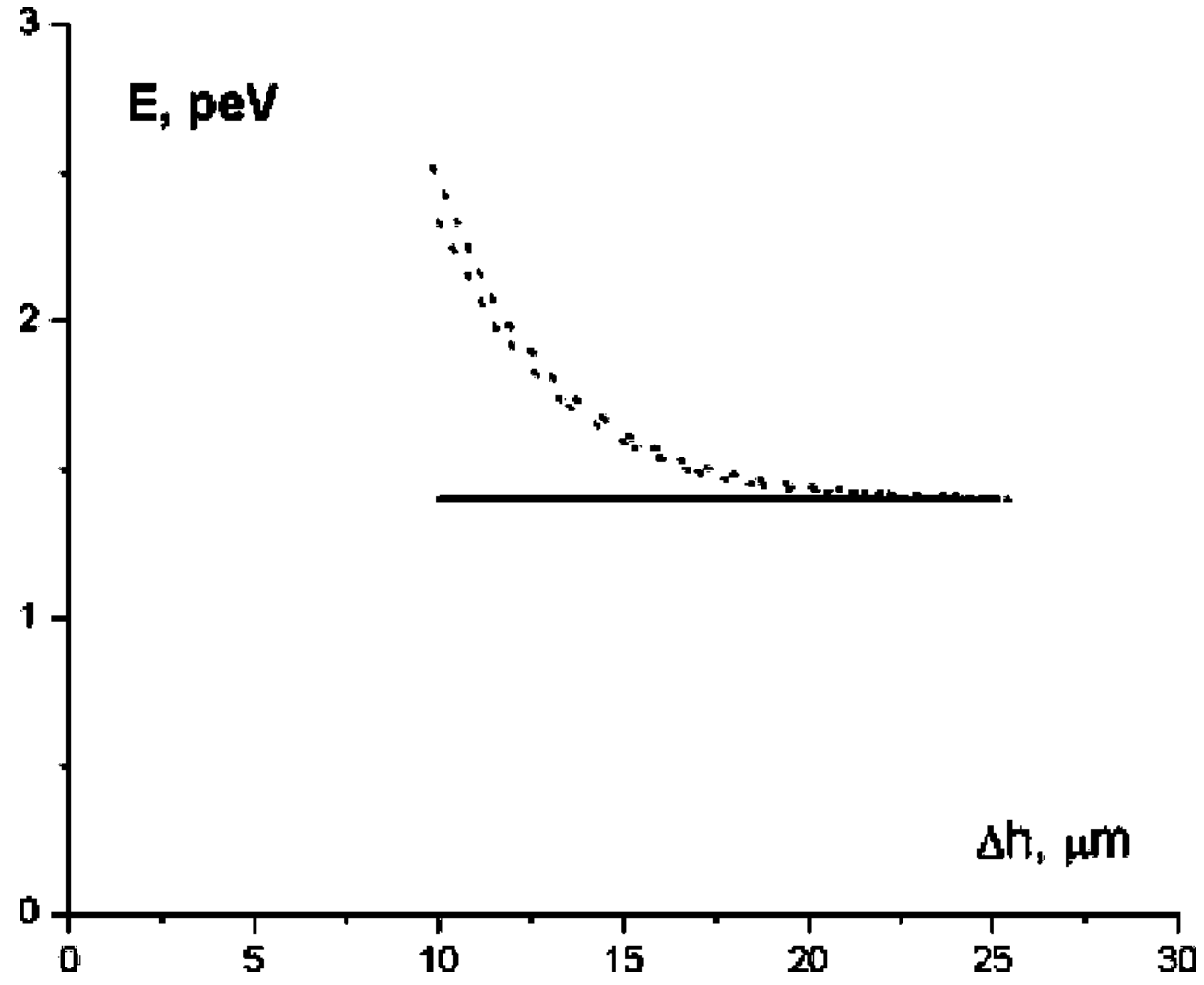

FIG. 7. Energy of the lowest quantum state as a function of the absorber height. The absorber strength for these two curves differs by a factor of 3. Horizontal line corresponds to the exact value $E_1$.

In this case the neutron flux at the detector is proportional to the number of "opened" levels. Thus

$$N(\Delta h)=\alpha\cdot\left[\sqrt{\frac{8\cdot g\cdot\Delta h^3}{9}}\cdot\frac{m}{\pi\cdot\hbar}-\frac{3}{4}\right], \qquad (8)$$

where the brackets [ ] mean the entire value, and $\alpha$ is a normalization coefficient. Obviously, the formula (8) reproduces the classical behavior (5) at large values of $\Delta h$. This steplike function can be written using the Heaviside function

$$\theta(x)=\begin{cases}1, & x>0,\\ 0, & x\leqslant 0,\end{cases} \qquad (9)$$

as follows:

$$N(\Delta h)=\alpha\cdot\sum_{n=0}^{\infty}\theta(\Delta h-h_n), \qquad (10)$$

where $h_n$ is the "height" of the $n$th level [see Eq. (1)]:

$$h_n=\sqrt[3]{\frac{9}{8\cdot g}\cdot\left[\left(n-\frac{1}{4}\right)\cdot\frac{\pi\cdot\hbar}{m}\right]^2}. \qquad (11)$$

Actually the absorber cuts the neutron spectrum at the energy slightly different from $m\cdot g\cdot\Delta h$ since the overlap of the neutron wave function and the absorber potential has to be sufficient for the absorption of a neutron. Therefore the values $h_n$ are shifted upwards by a small value of about $\delta=3$ $\mu$m (which is a constant since the corresponding neutron wave function shape is the same for all levels):

$$h_n'=h_n+\delta. \qquad (12)$$

The value $\delta=3$ $\mu$m can be estimated but it follows also from fitting the experimental data. The resulting function has to be "smoothened" by its convolution with a Gaussian resolution function with a dispersion of more than at least $d=2$ $\mu$m. This accounts for the experimental uncertainties (precision of the absorber height setting, inclination of the absorber, drifts of the slit size during the measurement, etc.):

$$N(\Delta h)=\frac{\alpha}{2}\cdot\sum_{n=1}^{\infty}\left[1+\mathrm{erf}\left(\frac{\Delta h-h_n'}{\sqrt{2}\cdot d}\right)\right]. \qquad (13)$$

Here erf() is the standard error function. Formula (13) describes well the experimental data at large heights; however, we need to introduce different occupation numbers $\nu_n$ for the few lowest levels:

$$N_{\mathrm{QM}}(\Delta h)=\frac{\alpha}{2}\cdot\sum_{n=1}^{\infty}\nu_n\cdot\left[1+\mathrm{erf}\left(\frac{\Delta h-h_n'}{\sqrt{2}\cdot d}\right)\right]. \qquad (14)$$

The formula (14) for $N_{\mathrm{QM}}(\Delta h)$ fits all experimental data as is shown in Fig. 5(c). The occupation numbers for fourth and higher levels are equal. The lowest level is significantly underpopulated; the third level is overpopulated. The reason for such behavior of the level populations has yet to be understood. It might be partly due to a possible small step between the two bottom mirrors, which can influence the level populations [5,24]. The results of full quantum-mechanical treatment do not differ qualitatively from the present simple description, but provide "smoother" curves.

TABLE I. The flux of neutrons, which penetrate between the bottom mirror and the absorber, is presented as a function of the absorber height. The velocity intervals are chosen such that the count rates are equal at large absorber heights.

| Absorber height ($\mu$m) | Count rate, $s^{-1}$, average horizontal velocity ~6 m/s (1) "Peak" discrimination window (2) "Counting of all events" | Count rate, $s^{-1}$, average horizontal velocity ~10 m/s (1) "Peak" discrimination window (2) "Counting of all events" |
|---|---|---|
| 19.5 | $0.0065 \pm 0.0003$ | $0.0060 \pm 0.0003$ |
| | $0.0113 \pm 0.0004$ | $0.0104 \pm 0.0004$ |
| 30.4 | $0.0135 \pm 0.0007$ | $0.0144 \pm 0.0008$ |
| | $0.0206 \pm 0.0009$ | $0.0219 \pm 0.0010$ |
| 40.7 | $0.0274 \pm 0.0010$ | $0.0265 \pm 0.0012$ |
| | $0.0411 \pm 0.0013$ | $0.0402 \pm 0.0014$ |

As evident from Fig. 5(c), the measured data, in particular, at $\Delta h < 20$ $\mu$m, do indeed strongly contradict the classical dependence $N_{CL}(\Delta h)$. They are in a good agreement with the quantum-mechanical expectations $N_{QM,0}(\Delta h)$ or $N_{QM}(\Delta h)$.

The count rate of neutrons, which penetrate between the bottom mirror and the absorber, was measured in two different wavelength ranges. The horizontal velocity component was selected and measured using the entrance collimator with two titanium plates. In one measurement the lower plate was in its lowest position and one half of the total intensity was shadowed with the upper plate. In the second measurement the upper plate was moved to its highest position and one half of the total intensity was shadowed with the lower plate. So the neutron spectrum was divided into two equal-intensity parts and their average horizontal velocity components were ~6 m/s and ~10 m/s, respectively.

The neutron fluxes in these intervals, as a function of the absorber height, are presented in Table I. Table I shows that the count rates do not depend on the horizontal neutron velocity component and correspondingly on the neutron wavelength. Only the vertical motion component is significant.

The efficiency of the absorber was measured. For such a test it was placed to the bottom and the 10 cm mirror was placed to the top (inverse geometry). Such an assembly is not transparent for neutrons if the efficiency of the absorber is 100%. The small penetration of neutrons through such an assembly provides an estimation for the absorber efficiency [see Fig. 5(b)]. The efficiency was found to be >95% for a 120 $\mu$m slit size and >98% for a 60 $\mu$m slit size [see Fig. 5(b)]. For the distance between two consequent collisions in the classic approximation [Eq. (2)], one gets an average absorber efficiency of >90% per collision. This estimation corresponds to the quality of the absorber, which is by far sufficient enough to have no significant related false effects in the transmission measurement. The excess of neutrons in the measurement with the absorber on the top compared with that with the absorber on the bottom (by a factor of ~20 at 120 $\mu$m and by a factor of ~50 at 60 $\mu$m slit sizes) is due to the confinement of neutrons in the gravitational field. Without gravity, these two measurements (in normal and reverse geometries) would give the same results [see Fig. 5(b)].

## V. ERROR ANALYSIS

Let us analyze the reliability of the experimental results and consider systematic errors, which could influence their interpretation, concerning the existence of the lowest quantum state. Any beam-related background is negligible when the absorber height is lower than ~15 $\mu$m (see Fig. 5).

The experimental results are well reproducible: the "non-transparency" at small absorber height was measured 11 times using different mirror configurations.

The precision (see above) of mirrors, absorber, and collimator is sufficient to clearly rule out any problem with their shape. X-ray-based studies of the roughness of the mirrors confirmed that also the probability of nonspecular reflections of neutrons should be smaller than a few percent (for details, see Ref. [24]). This is an important issue, because one can imagine a scenario, in which the transmission of neutrons can be suppressed due to a highly imperfect mirror. One can estimate the number of reflections at the bottom mirror in the classical approximation [using Eq. (2)] and take an estimated suppression factor at small absorber heights from the data shown in Fig. 5(c). This shows that the observed suppression of transmitted neutrons at small absorber heights could only be caused by multiple "reflections" of neutrons from a non-perfect mirror if the probability for nonspecular reflections was as high as ~60%, which contradicts the results of the x-ray measurements. Besides that, the neutron transmission would in this case depend strongly on the horizontal component of neutron velocity, which contradicts the observations (see Table I).

The installation was placed inside a $\mu$-metal shield and magnetic field gradients (which would be equivalent to a "curvature" of the mirror's surface) were measured to be sufficiently low.

In order to avoid false suppression of the neutron transmission through the slit between the mirror and the absorber,

caused by neutron diffraction at the front edge of the bottom mirror, we shaped a sufficiently broad angular distribution of the initial neutron beam. Thus the diffraction angle for neutrons at the lowest quantum state ($\lambda/h_1 \sim 4\times10^{-3}$, where $\lambda$ is the neutron wavelength) is much smaller than the angular divergence in the incoming neutron beam ($\sim 4\times10^{-2}$). Besides that, sizeable diffraction effects would depend strongly on the neutron wavelength, which is defined by the horizontal velocity component. This clearly contradicts the observations (see Table I).

There was no false effect caused by an "offset" in the height setting procedure. The maximal resulting error in the absorber height is a few micrometers if the height is $<100$ $\mu$m. It is not necessarily negligible at intermediate to large heights because of the error accumulation after many changes of the absorber height, and also because of poor control of slow drifts of the leg's height or the inclinometer electronics. In further experiments a direct absolute measurement using capacitors replaces this system. However, the reported measurement is free of noticeable systematic errors at a small absorber height. Many tests of reliability of the absorber positioning were carried out during the measurement. The height setting was controlled using a precise mechanical device (a comparator with the precision of $\sim 1$ $\mu$m). The absorber was raised and moved back to zero height. This is easy to control by inclinometers because the absorber then touches the bottom mirror. Reproducibility and stability of neutron data were permanently controlled. In some measurements we replaced the piezopositioning system by foil spacers of known thickness. Every test gave a positioning precision of at least a few micrometers at a small absorber height. The observation of transparency for visible light through the 15 $\mu$m slit between the bottom mirror and the absorber confirms also the absence of any sizeable offset in the absorber positioning. The neutron count rate through this slit was zero (see Fig. 5).

On the basis of the considerations presented we conclude that the slit between the bottom mirror and the absorber at the absorber height of $<15$ $\mu$m is not transparent for neutrons due to the quantum states of neutrons in the gravitational field.

## VI. CONCLUSION AND FURTHER STEPS

The lowest stationary quantum state of neutrons in the Earth's gravitational field was identified for the first time. We measured the transmission of neutrons between a long horizontal mirror on the bottom and an absorber on the top. We found that such an assembly is not transparent for neutrons if the absorber height is smaller than the spatial width of the lowest quantum state.

A clear observation of higher quantum states is a more difficult task than that for the lowest quantum state: In order to detect neutron flux showing that the first state is populated, one has to compare an almost zero signal (low background) with a nonzero signal. On the other hand, in order to resolve higher levels, one has to compare two nonzero and quite close signals. Besides that, the first step is the highest one, and the step height decreases quite rapidly with the state number. Therefore, even if step sharpness would be equal for all levels, the experimental separation of two neighboring levels is more difficult for higher levels. However, both the distance between levels as well as the experimental precision decreases with increasing level number. These reasons make the measurement of higher quantum levels more difficult. There are two main difficulties: The first one is the quality of the experimental setup itself (for instance, the accuracy of the absorber positioning, properties of the neutron beam, etc.), which can be, in principle, improved. The second one follows from quantum mechanics (the absorption of neutrons by the absorber is defined by an overlap of relatively smooth neutron wave function in the quantum state with the absorber profile) and cannot be easily changed within the setup and the method used here. The total uncertainty can be estimated from Fig. 5(c).

In the near future, we hope to improve significantly the experimental resolving power. The neutron flux could also be increased by optimization of neutron transport in front of the experimental setup. The accuracy of mechanical positioning of the optical elements will be improved such that the corresponding uncertainties would be largely reduced. These will allow to measure directly the transmission dependence, defined by quantum-mechanical limitations only.

In future generation experiments we would like to measure resonance transitions between different quantum states. The resonance method of measurement should provide ultimately values of the energies of quantum states, and therefore allow by far more precise results. Those could be used for different kinds of experiments, such as for the search for nonzero neutron charge or the search for extra fundamental interactions at small distances predicted in some modern quantum field theories (see, for example, Refs. [25,26], and references therein).

## ACKNOWLEDGMENTS

The authors would like to thank K. Ben-Saidane, D. Berruyer, Th. Brenner, J. Butterworth, D. Dubbers, E. Engel, P. Geltenbort, N. Havercamp, A. J. Leadbetter, A. Letsch, D. Mund, B. G. Peskov, S. V. Pinaev, R. Rusnyak, and M. Zimer for their help and for fruitful discussions. The present experimental program is partly supported by INTAS grant 99-705 and BMBF 06HD953. We are extremely grateful to all our colleagues who were interested in this research and contributed to its development.